\documentclass[twocolumn,amsmath,amssymb]{revtex4}

\usepackage{graphicx}
\usepackage{dcolumn}
\usepackage{bm}
\newcommand{\be}{\begin{equation}}
\newcommand{\ee}{\end{equation}}
\newcommand{\bea}{\begin{eqnarray}}
\newcommand{\eea}{\end{eqnarray}}

\begin{document}
\title{Einstein Gravity on the codimension 2 brane?}
\author{Paul Bostock$^1$}
\author{Ruth Gregory$^{1,2}$}
\author{Ignacio Navarro$^2$}
\author{Jose Santiago$^2$}

\affiliation{~$^1$Centre for Particle Theory, Department
of Mathematical Sciences\\
~$^2$Institute for Particle Physics Phenomenology, Department of Physics\\
University of Durham, DH1 3LE Durham, UK}

\begin{abstract}
We look at general braneworlds in six-dimensional
Einstein-Gauss-Bonnet gravity. We find the general
matching conditions for the Einstein-Gauss-Bonnet braneworld,
which remarkably turn out to give precisely the four-dimensional
Einstein equations for the induced metric and matter on the brane,
even when the extra dimensions are non-compact and have infinite
volume. We also show that relaxing regularity of
the curvature in the vicinity of the brane, or alternatively
having a finite width brane, gives rise to an additional
possible correction to the Einstein equations, which contains
information on the brane's embedding in the bulk and cannot
be determined from knowledge of the braneworld alone. We comment
on the advantages and disadvantages of each possibility,
and the relevance of these results regarding a possible solution
of the cosmological constant problem.

\vspace{0.3cm} \noindent PACS:11.25.Mj, 98.80.Jk \hspace{6.4cm}
IPPP/03/70, DCPT/03/49.
\end{abstract}


\maketitle

The braneworld paradigm, or the idea that our
universe might be a slice of some higher dimensional
spacetime, has proved a compelling alternative to
standard Kaluza-Klein (KK) methods of having more
than four dimensions.
Briefly, in contrast to KK {\it compactifications},
which have small and compact extra dimensions,
braneworlds can have large, even non-compact,
extra dimensions which have potentially important
experimental consequences \cite{ADD,RS}.
We do not directly see the extra dimensions since
we are confined to our braneworld, rather, their
presence is felt via short-scale corrections to
Newton's law, in some cases large scale
modifications of gravity, and as a means of generating
the hierarchy between the weak and Planck scales.
Although being confined to a slice in spacetime might
seem odd, such confinement is in fact a common occurance.
The early braneworld scenarios~\cite{EBW} for example
used zero modes on topological defects and of course
in string theory we have confinement of gauge
theories on D-branes.

Formally, the braneworld is a submanifold of the
spacetime manifold, and can have any number of
{\it co}-dimensions -- the number of extra dimensions -- up
to a maximum of 6/7 for string/M-theory.
By far the best investigated and understood braneworld
scenario is the codimension 1 case, or a toy
5-dimensional example motivated by the Horava-Witten
compactification of M-theory \cite{HW}.
This range of models, based on the seminal work of
Randall and Sundrum (RS) \cite{RS}, has all the features one
requires: Einstein gravity at some scale with
calculable modifications, well-defined cosmology
asymptoting standard cosmology at late times, and
has the additional allure of exhibiting directly
aspects of the adS/CFT correspondance.

Far less well understood are higher codimension braneworlds.
Although the pioneering work on resolving the hierarchy
problem took place within the context of higher codimension,
empirical models lack the gravitational consistency of the
RS scenarios. Attempts to include self-gravity have met
with some success in codimension 2 \cite{S,GV,C2R}, but for
codimension three or higher, the situation seems to
be more problematic \cite{CEG,GRSh}.

Codimension 2 brane worlds offer also some interesting properties 
that can be exploited to attack the cosmological constant problem \cite{C2R}, but one drawback is that, in contrast to
codimension 1, we appear to be very restricted in our
allowed brane energy-momenta. Typically, a brane in its
ground state has a very special energy-momentum tensor,
which is isotropic and has the property that Energy=Tension.
If we wish to have any matter on the brane, then we
have to have a varying energy density and varying tension.
However, as pointed out by Cline et\ al.\ \cite{CDGV}\ in the case
of cosmological branes, this is inconsistent with some
basic minimal assumptions about the nature of the braneworld.
Essentially, it causes singularities in the metric
around the braneworld which necessitates the introduction
of a cut-off and hence introduces questions of model
dependence.

In this letter we
suggest that the solution to the apparent
sterility of codimension 2 braneworlds might lie in the
Gauss-Bonnet term. This is a term that can be added to the action in
$D> 4$ (it is a topological invariant in 4D) that is quadratic in the
curvature tensor but has the well known property that the equations of
motion derived from it remain second order differential equations for
the metric. In fact, since ${\cal O}(R^2)$
corrections to the Einstein-Hilbert Lagrangian do arise in 
the low energy
limit of string theory, the inclusion of this type of
term could be regarded as mandatory if one wants to embed any
braneworld solution into string/M-theory.
Fluctuations around a flat background for this model were studied 
in \cite{Corradini:2001qv},  
and the conclusions obtained are compatible with the ones presented 
in this Letter at the linearized level.

In trying to derive effective Einstein equations on
the brane, it is worth comparing and contrasting with
codimension 1. Recall that for codimension 1 there
is a single normal to the braneworld, hence a single
direction from the braneworld. For a general submanifold
of codimension 2, there are now {\it two} normals, and
for a regular submanifold we again have a well-defined
coordinate patch around it defined by the Gauss-Codazzi
formalism. (This method was used to derive effective
actions of topological defects \cite{EFA}.)
The problem with trying to apply Gauss-Codazzi in our
case is that it requires some minimal regularity
of the metric near the braneworld, and this is no
longer the case for an infinitesimal braneworld
in codimension 2 -- the situation
is even worse for codimension 3 and higher!
Briefly, there is no well-defined ``thin braneworld''
limit for the Einstein equations \cite{GT}, or alternatively,
for the conical deficit, it is not possible to
put two normals at the location of the deficit
which have a well defined inner product -- it depends
on whether you measure the outer or inner angle.
In order to derive gravity on the brane therefore,
we instead use a coordinate system which is defined
in the vicinity of the braneworld, and in which the
effect of the brane formally appears as a delta-function.

We assume that our braneworld has a nonsingular metric,
$\hat{g}_{\mu\nu} (x^\mu)$, which is continuous in the vicinity
of the braneworld. The coordinates $x^\mu$ label the
braneworld directions, and we will use greek indices to indicate
braneworld coordinates. We now take the set of points
at a fixed proper distance from a particular $x^\mu$ on
the brane, this will have topology $S^1$, and we label
these points by $x^\mu$, their proper distance, $r$, from
$x^\mu$, and an angle $\theta$, which without loss of
generality we will take to have the standard periodicity
of $2\pi$. This method provides a full coordinatisation of
spacetime in the vicinity of the braneworld, and will
be unique within the radii of curvature of the braneworld.
There are two remaining issues. One is that there is of
course some ambiguity in the labelling of $\theta$, which
is equivalent to the choice of connection on the normal
bundle of the braneworld, however, for simplicity we will
assume that $\theta$ is chosen to make this connection
vanish (in particular, this means we assume that the
braneworld is not self intersecting). The second issue relates
to the form of the bulk spacetime metric, which we will
now assume has axial symmetry, {\it i.e.}, $\partial_\theta$
is a Killing vector. This Ansatz simplifies the bulk metric,
and it is analogous to the assumption of Z$_2$-symmetry
in the codimension 1 scenarios. The metric therefore can
be seen in these coordinates to take the general form:
\be
d s^2= g_{\mu\nu}(x,r)dx^\mu dx^\nu -
L^2(x,r) d \theta^2 - d r^2\label{metric}.
\ee

In order to obtain the braneworld equations, we now expand the
metric around the brane:
\be
L(x,r)= \beta(x) r + O(r^2)
\ee
etc. For values of $\beta \neq 1$ we have a conical singularity
at $r=0$, which is interpreted as being due to a delta-function
braneworld source. Strictly speaking, at least in Einstein gravity,
we cannot talk of a delta-function source in terms of a
zero-thickness limit of finite sources \cite{GT}, rather, we
deduce the existence of the delta-function in the Riemann tensor
from the holonomy of a parallely transported vector around the
source. However, as the equations of motion make perfect sense
with the delta-function being encoded in a notional
discontinuity of the radial derivatives of the metric at $r=0$,
we follow the standard procedure in this paper of defining
$L'(x,0)=1$, $g'_{\mu\nu}(x,0)=0$, in order to give rise to the
required distributional behaviour of the curvature in the
gravitational equations (a prime denotes derivative with respect to $r$).

Therefore, for a general braneworld, the problem we wish to
solve is that of finding gravitating solutions that include
the effect of a general brane energy-momentum tensor
\be
T_{MN}=\left(\begin{array} {cc}
\hat{T}_{\mu \nu}(x)\frac{\delta(r)}{2\pi L} & 0\\
0 &  0 \end{array}\right)
\ee
(upper case latin indices run over all the dimensions). In
particular, we will be interested in the relation between the
4D induced metric on the brane, $g_{\mu \nu}(x,0)=
\hat{g}_{\mu \nu}(x)$, and the brane energy momentum tensor,
$\hat{T}_{\mu \nu}(x)$. It is this relation which
determines the nature of the gravitational interactions that a
``brane observer'' would measure.

Our starting point is the Einstein-Gauss-Bonnet (EGB) equation
\be
M_\ast^4 \left ( G_{MN}+H_{MN} \right ) = T_{MN} + S_{MN},
\label{eq.GB}
\ee
\noindent where
\begin{equation}
G_{MN}=R_{MN}-\frac{1}{2}g_{MN}R,
\end{equation}
and the Gauss-Bonnet contribution is given by
\bea
H_{MN}=
\alpha\Big[ & \frac{1}{2}g_{MN}(R^2-4 R^{PQ}R_{PQ}+R^{PQST}R_{PQST})
\nonumber \\
-&2 R R_{MN}+4 R_{MP}R_N^{\phantom{N}P}+4R^K_{\phantom{K}MPN}
R_K^{\phantom{K}P}
\nonumber \\
&-2R_{MQSP}R_N^{\phantom{N}QSP}\Big],
\eea
with $\alpha$ a parameter with dimensions of $(mass)^{-2}$.
$S_{MN}$ is the bulk energy momentum tensor, which we
will not specify here, other than to assume that it has no
delta-function contributions.

If equation (\ref{eq.GB}) is to be satisfied there must be a singular
contribution to the LHS of this equation with the
structure $\sim \frac{\delta(r)}{L}$.  As we have already
discussed, such a contribution can arise from terms
that contain:
\bea
\frac{L^{\prime\prime}}{L} &=& - (1-\beta) \frac{\delta(r)}{L}
+\;\;({\rm non-singular\;\; part}),\\
\frac{\partial_r^2 g_{\mu \nu}}{L} &=& \partial_r g_{\mu \nu}
\frac{\delta(r)}{L}+\;\;({\rm non-singular\;\; part}).\label{deltag}
\eea
In Einstein gravity, these latter terms are zero.
However, since they could in principle be nonzero,
we will retain them from now.

We must therefore set the delta-function contribution equal
to the brane energy-momentum tensor in order to solve the
equations of motion. After some calculation, one obtains
that the only singular part of the LHS of equation (\ref{eq.GB}) 
lies in the $\mu, \nu$ directions and is:
\be
-\frac{L^{\prime\prime}}{L}
\left[ g_{\mu\nu}+ 4 \alpha \left( R_{\mu\nu}(g)-\frac{1}{2}
g_{\mu\nu} R(g) \right) \right ]
+ \frac{\alpha}{L} \partial_r \left(
L^\prime \;W_{\mu\nu} \right),
\ee
where $W_{\mu\nu}$ is defined as the following combination of
first derivatives of the 4-dimensional metric
\bea
W_{\mu\nu}&=&
\,
g^{\lambda\sigma}\partial_r g_{\mu\lambda}\partial_r g_{\nu\sigma}
-
g^{\lambda\sigma}\partial_r g_{\lambda\sigma}\partial_r g_{\mu\nu}
\nonumber \\
&+&\frac{1}{2} g_{\mu\nu}
\left [ (g^{\lambda\sigma}\partial_r g_{\lambda\sigma})^2
- g^{\lambda\sigma}g^{\delta\rho}
\partial_r g_{\lambda\delta}\partial_r g_{\sigma\rho}\right ] .
\eea

We can now use the properties
\bea
-\frac{L^{\prime\prime}}{L} &=& (1-\beta) \frac{\delta(r)}{L}
+\ldots, \\
\frac{\partial_r(L^\prime \; W_{\mu\nu})}{L}&=&
\beta \;  W_{\mu\nu}|_{r=0^+}  \frac{ \delta(r) }{L}+ \ldots,
\eea
to obtain the matching condition by equating the $\frac{\delta(r)}{L}$ terms of equation (\ref{eq.GB}). This yields
\begin{equation}
2\pi(1-\beta) M_\ast^4 \big[\hat{g}_{\mu\nu}
+ 4 \alpha \hat{G}_{\mu\nu} + \alpha \frac{\beta}{1-\beta}
\hat{W}_{\mu\nu} \big] = \hat{T}_{\mu\nu},
\label{grav.br}
\end{equation}
where $\hat{G}_{\mu\nu}$ is the 4D Einstein tensor for the induced metric, $\hat{g}_{\mu\nu}$, and $\hat{W}_{\mu\nu}\equiv W_{\mu\nu}|_{r=0^+}$.

This is our main result: the gravitational equations
of a braneworld observer are the Einstein equations plus
an extra Weyl-term, $\hat{W}_{\mu\nu}$, which depends on the
bulk solution. This term is reminiscent of the Weyl term in
the codimension 1 braneworlds \cite{SMS}, which gives rise to
the corrections to the Einstein equations on the brane. Roughly
speaking, the braneworld equation is obtained by
taking the components of the full Einstein equations parallel
to the brane, with the perpendicular components giving some
information on the nature of the Weyl term. Depending on the
symmetries present, in some cases (cosmology being the most
physically interesting) we can completely determine
the bulk metric, and hence these Weyl corrections.
For codimension 2, the perpendicular components of the bulk equations
do lead to constraints as we discuss presently, however these now
no longer fix the bulk metric exactly, not even for the highly
symmetric and special case of braneworld cosmology with Einstein
gravity in the bulk. Let us now investigate the consequences of
(\ref{grav.br}), in particular, the consistency of the extra
Weyl term, which arose as a result of allowing a discontinuity
in the derivative of the parallel braneworld metric.

A natural first check is to take the $\alpha \rightarrow 0$
limit to recover the Einstein case. Then equation
(\ref{grav.br}) reduces to
\be
2\pi(1-\beta) M_\ast^4 \hat{g}_{\mu\nu}=\hat{T}_{\mu\nu}.
\ee
Although this looks like it is not possible to satisfy this
matching condition unless the brane energy momentum tensor
is proportional to the induced metric,
in fact we have not yet determined whether $\beta$ is a constant.
A non-constant $\beta$ would correspond to a varying deficit
angle, and is not determined by the braneworld equations alone.
We must supplement the braneworld equations with the bulk equations
normal to the braneworld, and since we wish to make as few
assumptions as possible about the bulk in this letter, we will
simply look at the divergent ${\cal O}(1/r)$ terms in the
Einstein equations near the brane, as these cannot be cancelled
by any regular bulk $S_{MN}$. These leading terms for
the $(\mu, \nu)$, $(r,r)$ and $(\mu,r)$ components give
\bea
g_{\mu \nu}\frac{[L^{\prime\prime}]}{L} - \frac{L^\prime}{2L}\left[
\partial_r g_{\mu \nu}- g_{\mu \nu}g^{\rho \sigma}\partial_r g_{\rho
\sigma}\right]=0,&\nonumber\\
\frac{L^\prime}{2L}g^{\rho \sigma}\partial_r g_{\rho \sigma}=0
\;\;\; ,  \;\;\;
\frac{\partial_\mu L^\prime}{L}=0,&
\eea
where $[L^{\prime\prime}]$ stands here for the smooth part of the second
derivative as we approach the brane.
We now see directly that $\beta$ must indeed be constant,
and that $\partial_r^2 L|_{r=0^+}=0$ and $\partial_r
g_{\mu\nu}|_{r=0^+}=0$. We now confirm the observation of Cline
et\ al.\ \cite{CDGV}, that Einstein codimension 2 braneworlds
must have an energy momentum proportional to their induced
metric, and their gravitational effect is to produce a conical
deficit in the bulk spacetime.

In Gauss-Bonnet gravity however, the situation is not so simple,
since all these equations get corrections proportional to $\alpha$
and one cannot rule out the existence of solutions with
$\hat{W}_{\mu \nu} \neq0$.
The ${\cal O}(1/r)$ terms in the $(\mu,r)$ components of the
EGB equations for example are
\bea
&-&g^{\nu\sigma}\frac{\partial_\sigma L^{\prime}}{L}
\Big[ g_{\mu\nu}+ 4 \alpha \Big( R_{\mu\nu}(g)-\frac{1}{2}
g_{\mu\nu} R(g) \Big) -\alpha\; W_{\mu\nu} \Big]
\nonumber \\
&+&2\alpha \frac{L^\prime}{L} g^{\nu\sigma}
\Big[ \partial_r g_{\mu\nu}\; R^\rho_{\; \sigma\rho r}
-\partial_r g_{\nu\sigma}\; R^\rho_{\; \mu\rho r}
\Big]=0,\label{1/r:mu_r}
\eea
with similar constraints from the $\mathcal{O}(1/r)$ terms of the
$(\mu,\nu)$ and $(r,r)$ equations (though these are somewhat
more complicated and not particularly illuminating).
In this case we find that in general no simple
restriction can be placed on the solution, and in particular the
deficit angle $\beta$ need no longer be constant.

However, it is important to note that
some components of the Ricci curvature tensor (and scalar)
are now divergent once we allow $\partial_r g_{\mu\nu}|_{0^+}\neq0$.
For example
\be
R_{\mu\nu}= \frac{1}{2} \frac{L^\prime}{L} \partial_r g_{\mu\nu}
+ \ldots = \frac{\partial_r g_{\mu\nu}}{2r} + \mathcal{O}(1),
\ee
near the brane. In a realistic situation, we could argue
that a brane would have finite width, which could act as a
cut-off for the curvature, hence all our results would still
be valid provided this cut-off is sufficiently large so that
the curvature is still small compared to $M_\ast^2$, the
six-dimensional Planck mass squared.
In this smooth case, we {\it can} use the Gauss-Codazzi
formalism and the $\theta$-independence of the metric to write
\be
W_{\mu\nu} = K_{_i\mu}^\lambda K_{_i\nu\lambda}
-K_{_i} K_{_i\mu\nu} + \frac{1}{2} g_{\mu\nu}
\left [ K_{_i}^2 - K_{_i\lambda\sigma}^2 \right ],
\ee
where $K_{_i\mu\nu}$ are the two extrinsic curvatures
($i=1,2$) for each of the two normals. We therefore have
the interpretation of $W_{\mu\nu}$ as a
geometric correction to the Einstein
tensor due to the embedding of the braneworld in the bulk
geometry.
The interpretation is then that the Einstein equations
acquire additional embedding terms which unfortunately cannot
be deduced from the braneworld geometry alone.

The physical relevance of terms which lead to divergent
curvatures and hence tidal forces in the vicinity of the
braneworld is however questionable. If $M_\ast$ is of
order the (inverse) brane width,
or if we wish to have a truly infinitesimal brane, then we
are forced to conclude that for consistency we cannot stop
at the GB curvature corrections, but must include all
higher order curvature corrections thus entering a
non-perturbative regime of which we can say nothing.
We are therefore forced to impose $\partial_r g_{\mu\nu}=0$,
and equation (\ref{1/r:mu_r}) tells us that the
deficit angle $\beta$ is again constant and the equation for the
induced metric (\ref{grav.br}) remarkably takes the form of purely
four-dimensional Einstein gravity
\be
\hat{G}_{\mu \nu}=
\frac{1}{8\pi(1-\beta)\alpha M_\ast^4}
\hat{T}_{\mu \nu}
- \frac{1}{4 \alpha} \hat{g}_{\mu \nu}.
\ee
We can read off our 4
dimensional Planck mass as
\be
M_\mathrm{Pl}^2= 8\pi(1-\beta)\alpha M_\ast^4,
\ee
and we note the presence of
an effective 4 dimensional cosmological constant
\be
\Lambda_4 = T_0-2\pi(1-\beta)M_\ast^4,
\ee
where $T_0$ is the bare brane tension:
\be
{\hat T}_{\mu\nu} = T_0\ {\hat g}_{\mu\nu} + \delta T_{\mu\nu}.
\label{split}
\ee
Of course the splitting of the energy-momentum tensor in
this manner is potentially arbitrary, however, for a cosmological
brane $\delta T_{\mu\nu} \to 0$ as $t \to \infty$, and we
can simply posit that  $\delta T_{\mu\nu} \to 0$ as either
$t$ or $|{\bf x}| \to \infty$ as being a necessary
requirement of a braneworld thus rendering (\ref{split}) unambiguous.

Interestingly, the Einstein relation between $\beta$ and the
brane tension: $T_0=2\pi (1-\beta)M^4_\ast$
no longer holds for GB gravity -- we can specify the
conical deficit and the brane tension independently, the only
caveat being that if the Einstein relation does not hold, then
we have an effective cosmological constant on the brane.

To sum up: we have found the equations governing the induced metric on
the brane for a codimension 2 braneworld. We have shown that adding
the Gauss-Bonnet term allows for a realistic gravity on an
infinitesimally thin brane
which remarkably turns out to be precisely four-dimensional Einstein
gravity {\it independent} of the precise bulk structure, the only
bulk dependence appearing via the constant deficit angle $\Delta$
in the definition of the 4-dimensional Planck mass
$M_{\rm Pl}^2 = 4\alpha \Delta M_\ast^4$. Since Einstein gravity
appears quite generically, our model provides a novel alternative
realization of the infinite extra dimensions idea of Dvali et\ al.\ \cite{Dvali:2000hr}. Indeed, we could modify our model by adding
braneworld Ricci terms (which can be motivated via finite width
corrections to the brane effective action \cite{EFA}), which would
give the same form of the braneworld gravity equations, and simply
renormalize the 4-dimensional Planck mass.

We also showed that it was possible to obtain a deviation from
Einstein gravity via a non-zero $\hat{W}_{\mu\nu}$. In turn,
this allows a variation of the bulk deficit angle and therefore
the effective brane cosmological constant. In this case, one
has to either perform a smooth regularization of the brane by
taking some finite width vortex model, or accept that the
infinitesimally thin braneworld has a non-perturbative regime
in the neighbourhood of the brane. Nevertheless it seems to be
a very appealing feature towards a possible solution of the
cosmological constant problem. One could envisage a situation in which
the system is in a non-perturvative phase in which the cosmological
constant can vary, and relax itself dynamically to a perturbative
state in which the induced gravity on the brane is four-dimensional
Einstein gravity and with a very small
cosmological constant (an infinite flat supersymmetric bulk might for
instance lead to this situation~\cite{Witten:2000zk}).
Due to the unbounded curvature near the brane when this situation is
violated it seems plausible that once the system reaches that
configuration it would prefer to remain there.


\begin{thebibliography}{9}

\bibitem{ADD}
N.~Arkani-Hamed, S.~Dimopoulos and G.~R.~Dvali,
Phys.\ Lett.\ B {\bf 429}, 263 (1998)
[arXiv:hep-ph/9803315],
Phys.\ Rev.\ D {\bf 59}, 086004 (1999)
[arXiv:hep-ph/9807344];
I.~Antoniadis, N.~Arkani-Hamed, S.~Dimopoulos and G.~R.~Dvali,
Phys.\ Lett.\ B {\bf 436}, 257 (1998)
[arXiv:hep-ph/9804398].


\bibitem{RS}
L.~Randall and R.~Sundrum,
Phys.\ Rev.\ Lett.\  {\bf 83}, 3370 (1999)
[arXiv:hep-ph/9905221],
Phys.\ Rev.\ Lett.\  {\bf 83}, 4690 (1999)
[arXiv:hep-th/9906064].

\bibitem{EBW}
V.~A.~Rubakov and M.~E.~Shaposhnikov,
Phys.\ Lett.\ B {\bf 125}, 139 (1983),
Phys.\ Lett.\ B {\bf 125}, 136 (1983);
K.~Akama,
Lect.\ Notes Phys.\  {\bf 176}, 267 (1982)
[arXiv:hep-th/0001113].

\bibitem{HW}
P.~Horava and E.~Witten,
Nucl.\ Phys.\ B {\bf 475}, 94 (1996)
[arXiv:hep-th/9603142].

\bibitem{S}
R.~Sundrum,
Phys.\ Rev.\ D {\bf 59}, 085010 (1999)
[arXiv:hep-ph/9807348].

\bibitem{GV}
A.~G.~Cohen and D.~B.~Kaplan,
Phys.\ Lett.\ B {\bf 470}, 52 (1999)
[arXiv:hep-th/9910132];
R.~Gregory,
Phys.\ Rev.\ Lett.\  {\bf 84}, 2564 (2000)
[arXiv:hep-th/9911015].

\bibitem{C2R}
J.~W.~Chen, M.~A.~Luty and E.~Ponton,
JHEP {\bf 0009} (2000) 012
[arXiv:hep-th/0003067];
S.~M.~Carroll and M.~M.~Guica,
arXiv:hep-th/0302067;
I.~Navarro,
JCAP {\bf 0309} (2003) 004
[arXiv:hep-th/0302129],
Class.\ Quant.\ Grav.\  {\bf 20}, 3603 (2003)
[arXiv:hep-th/0305014];
Y.~Aghababaie, C.~P.~Burgess, S.~L.~Parameswaran and F.~Quevedo,
arXiv:hep-th/0304256.

\bibitem{CEG}
C.~Charmousis, R.~Emparan and R.~Gregory,
JHEP {\bf 0105}, 026 (2001)
[arXiv:hep-th/0101198].

\bibitem{GRSh}
T.~Gherghetta, E.~Roessl and M.~E.~Shaposhnikov,
Phys.\ Lett.\ B {\bf 491}, 353 (2000)
[arXiv:hep-th/0006251].

\bibitem{CDGV}
J.~M.~Cline, J.~Descheneau, M.~Giovannini and J.~Vinet,
JHEP {\bf 0306}, 048 (2003)
[arXiv:hep-th/0304147].

\bibitem{Corradini:2001qv}
O.~Corradini, A.~Iglesias, Z.~Kakushadze and P.~Langfelder,
Phys.\ Lett.\ B {\bf 521} (2001) 96
[arXiv:hep-th/0108055].


\bibitem{EFA}
R.~Gregory,
Phys.\ Rev.\ D {\bf 43}, 520 (1991);

\bibitem{GT}
R.~Geroch and J.~Traschen,
Phys.\ Rev.\ D {\bf 36}, 1017 (1987).


\bibitem{SMS}
T.~Shiromizu, K.~i.~Maeda and M.~Sasaki,
Phys.\ Rev.\ D {\bf 62} (2000) 024012
[arXiv:gr-qc/9910076];


\bibitem{Dvali:2000hr}
G.~R.~Dvali, G.~Gabadadze and M.~Porrati,
Phys.\ Lett.\ B {\bf 485} (2000) 208
[arXiv:hep-th/0005016].


\bibitem{Witten:2000zk}
E.~Witten,
``The cosmological constant from the viewpoint of string theory,''
arXiv:hep-ph/0002297.

\end{thebibliography}
\end{document}